\documentclass[twocolumn,english,aps,prl,superscriptaddress,bibnotes,amsmath,amssymb]{revtex4-1}
\usepackage[T1]{fontenc}
\usepackage[latin9]{inputenc}
\usepackage{graphicx}

\makeatletter
\usepackage{dcolumn}
\usepackage{bm}
\usepackage{color}
\usepackage{soul}
\usepackage{hyperref}

\usepackage{lmodern}
\usepackage[T1]{fontenc}

\makeatother

\usepackage{babel}
\begin{document}

\title{Universal dynamics and controlled switching of dissipative Kerr solitons
in optical microresonators}

\author{M. Karpov}
\thanks{M.K., H.G. and E.L. contributed equally to this work}
\affiliation{École Polytechnique Fédérale de Lausanne (EPFL), CH-1015 Lausanne,
Switzerland}

\author{H. Guo}
\thanks{M.K., H.G. and E.L. contributed equally to this work}
\affiliation{École Polytechnique Fédérale de Lausanne (EPFL), CH-1015 Lausanne,
Switzerland}

\author{E. Lucas}
\thanks{M.K., H.G. and E.L. contributed equally to this work}
\affiliation{École Polytechnique Fédérale de Lausanne (EPFL), CH-1015 Lausanne,
Switzerland}

\author{A. Kordts}
\affiliation{École Polytechnique Fédérale de Lausanne (EPFL), CH-1015 Lausanne,
Switzerland}

\author{M.H.P. Pfeiffer}
\affiliation{École Polytechnique Fédérale de Lausanne (EPFL), CH-1015 Lausanne,
Switzerland}

\author{G. Lihachev}
\affiliation{Faculty of Physics, M.V. Lomonosov Moscow State University, 119991
Moscow, Russia}
\affiliation{Russian Quantum Center, Skolkovo 143025, Russia}

\author{V.E. Lobanov}
\affiliation{Russian Quantum Center, Skolkovo 143025, Russia}

\author{M.L. Gorodetsky}
\affiliation{Faculty of Physics, M.V. Lomonosov Moscow State University, 119991
Moscow, Russia}
\affiliation{Russian Quantum Center, Skolkovo 143025, Russia}

\author{T.J. Kippenberg}
\email[]{tobias.kippenberg@epfl.ch}
\affiliation{École Polytechnique Fédérale de Lausanne (EPFL), CH-1015 Lausanne,
Switzerland}

\date{\today}

\begin{abstract}
Dissipative temporal Kerr solitons in optical microresonators enable
to convert a continuous wave laser into a train of femtosecond pulses.
Of particular interest are single soliton states, whose $\mathrm{sech}^{2}$
spectral envelope provides a spectrally smooth and low noise optical
frequency comb, and that recently have been generated in crystalline,
silica, and silicon-nitride resonators. They constitute
sources that are unique in their ability to provide short femtosecond
pulses at microwave repetition rates. Likewise, they provide essential
elements to realize chip-scale, integrated frequency combs for time-keeping,
spectroscopy, navigation or telecommunications. However to date, the
dynamics of this class of solitons in microresonators remains largely
unexplored, and the reliable generation of single soliton states remains
challenging. Here, we study the dynamics of multiple soliton states
containing ${N}$ solitons and report the discovery of a novel, yet
simple mechanism which makes it possible to reduce deterministically
the number of solitons, one by one, i.e. ${N\! \to\! N\!-\!1\! \to\! \dots \!\to\! 1}$.
By applying weak phase modulation, we directly characterize the soliton state via a double-resonance response. The dynamical probing demonstrates that transitions occur in a predictable way, and thereby enables us to map experimentally the underlying multi-stability diagram of dissipative Kerr solitons.
These measurements reveal the ``lifted'' degeneracy of soliton states as a result of the power-dependent thermal shift of the
cavity resonance (i.e. the thermal nonlinearity). The experimental results are in agreement with theoretical and numerical analysis that incorporate the thermal nonlinearity. By studying two different microresonator platforms  (integrated $\mathrm{Si_{3}N_{4}}$  microresonators and crystalline $\mathrm{MgF_{2}}$
resonators) we confirm that these effects have a universal nature. Beyond elucidating the fundamental
dynamical properties of dissipative Kerr solitons the observed phenomena are also of practical relevance, providing a manipulation toolbox which enables to sequentially reduce, monitor and stabilize the number ${N}$ of solitons,  preventing it from decay. Achieving reliable single soliton operation
and stabilization in this manner in optical resonators is imperative
to applications.
\end{abstract}
\maketitle

\label{Introduction}

Microresonator frequency combs (Kerr combs) have opened a novel research
area at the interface of micro- and nano-photonics and frequency metrology
\cite{delhaye2007comb,Kippenberg2011microcombs}. Kerr combs are generated
in high-$Q$ millimeter- or micron-scale
resonators via parametric processes driven by continuous wave (CW) laser \cite{Kippenberg2004parametric,Savchenkov2004parametric}.
Kerr combs have attracted significant attention over past years due to unprecedented compactness, demonstrated octave-spanning operation
\cite{DelHaye2011octavecomb,Okawachi2011octavespan}, repetition rates
in the microwave domain ($>\!10~\mathrm{GHz}$), and ability to be operated in low noise regimes \cite{Herr2012universal,ferdous2011linebylineshape,Delhaye2013parametric}.
They promise chip-scale optical frequency combs connecting RF to optical domain that could make metrology ubiquitous, widely accessible beyond specialized metrology laboratories.
Recently, it has been demonstrated that Kerr combs can be operated in the regime of
temporal dissipative Kerr solitons (DKS) \cite{herr2014soliton,Brasch2014Cherenkov}.
DKS allow for fully coherent optical frequency combs
(soliton combs) that can be sufficiently broadband for self-referencing
via soliton induced Cherenkov radiation \cite{Brasch2014Cherenkov},
and provide access to stable ultrashort pulses of tunable duration
\cite{herr2014soliton,karpov2015raman} at microwave repetition rates
\cite{vahala2015soliton}. Of particular interest are \textit{single
soliton} states, that exhibit a spectrally smooth $\mathrm{sech^{2}}$
envelope. Such soliton based
frequency comb sources have a wide range of applications
including molecular spectroscopy \cite{ideguchi2013coherent}, coherent
data transmission \cite{pfeifle2014coherent,Pfeifle20tbs}, arbitrary waveform generation \cite{ferdous2011linebylineshape}, optical clocks \cite{Papp2014kerropticalclock} or astrophysics \cite{Kippenberg2011microcombs},
and more generally in applications where short pulse duration at microwave
repetition rate is desirable.

Originally discovered to spontaneously form in crystalline $\mathrm{MgF}_{2}$
resonators \cite{herr2014soliton} (and for the first time externally
induced in optical fiber cavities \cite{leo2010temporal}), DKS have been demonstrated in a variety of high-$Q$ resonator
platforms, ranging from silica wedge resonators \cite{vahala2015soliton},
to $\mathrm{Si_{3}N_{4}}$ photonic chips \cite{Brasch2014Cherenkov}
and compact crystalline resonators pumped via distributed feedback
lasers \cite{matsko2015soliton,Grudinin2015soliton}. Due to the recent
nature of these findings, the soliton formation process and its dynamics
remain to date largely unexplored. While solitons have been
reported in a number of platforms, the soliton generation procedures
in high-$Q$ microresonators are inherently stochastic (techniques
used in optical fiber cavities \cite{jang2015temporal} are technically
impractical due to much shorter round-trip time of microresonators).
While CW laser tuning and ``power kicking'' schemes were proposed \cite{Brasch2014Cherenkov,vahala2015soliton} for soliton
generation, these techniques presently do not allow to control the number
of solitons formed in the resonator. Another important question is the possibility
of deterministic manipulation of states with multiple solitons in
microresonators. Even though the states with various number of solitons
could be generated in optical microresonators, the transitions between
them take place stochastically via pairwise interactions of solitons
when the pump is tuned, and cannot be predicted so far. Due to these
effects, deterministic generation of the single soliton state still
represents an outstanding challenge. One more challenge is the non-destructive
monitoring of the soliton state. The soliton regime in microresonators
is fragile (though self-sustainable) and is not persistent against
significant thermal drifts and other external perturbations. The reported
passive lifetime of DKS achieves several hours in a stable laboratory
environment \cite{Brasch2014Cherenkov,herr2014soliton}, however,
no technique is known to enable feedback stabilized
control of soliton state, preventing it from decay.

\begin{figure*}
\includegraphics[width = 2 \columnwidth]{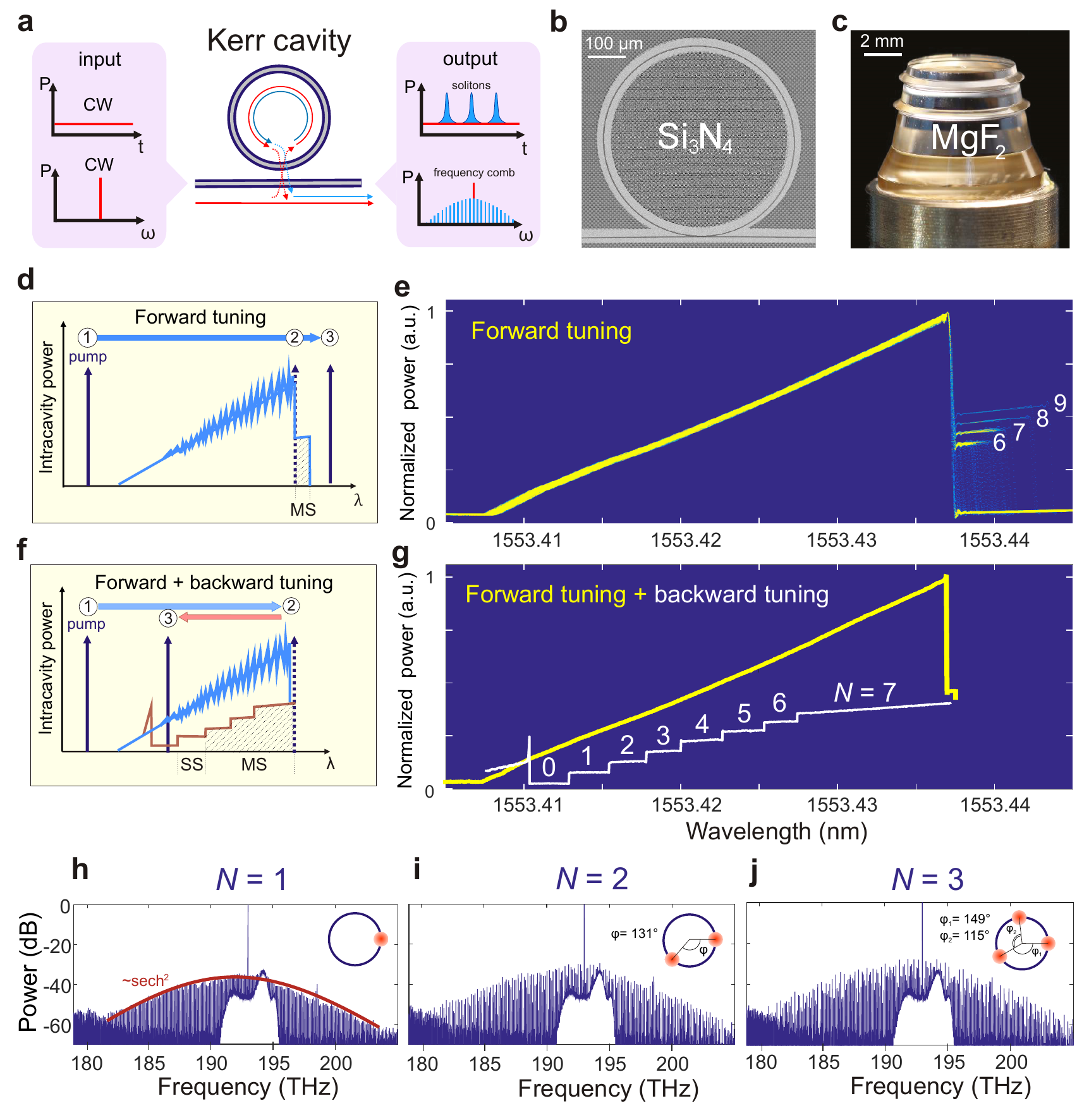}
\protect\caption{\textbf{Forward and backward tuning of the pump.} (a) The principle of microresonator frequency comb generation and the formation of dissipative Kerr solitons; (b) SEM image of the ${\mathrm{Si_{3}N_{4}}}$ on-chip microresonator with a free spectral range
(FSR) of $100~\mathrm{GHz}$; (c) Picture of the $\mathrm{MgF}_{2}$
crystalline resonator with FSR of $14~\mathrm{GHz}$; (d) Scheme
of the laser tuning method for the soliton generation in optical microresonators.
The pump laser is tuned over the resonance from short to long wavelengths
(forward tuning). Hatched region indicates the pump detuning range
of multiple solitons (MS); (e) Histogram plot of $200$ overlaid experimental
traces of the output comb light in the pump forward tuning over the resonance with the same
pump power and tuning speed, which reveals the formation of a predominant
multiple soliton state with ${N=6}$; The noise pattern in the forward detuning was not captured by the measurements due to the averaging in the photodetector; (f) Scheme of the laser backward
tuning. To initiate the sequence, the forward tuning is first applied,
and the pump is stopped in a multiple soliton state (which can be stable
by suitable choice of the laser tuning speed). In the second stage,
the pump is tuned back to short wavelengths, which leads to successive
soliton switching, ${N\! \to\! N\!-\!1\! \to\! \dots \!\to\! 1}$.
The MS area indicates the detuning range of multiple soliton states,
which is much larger compared to the forward tuning method. There
also exists the range of the single soliton state (SS); (g) Experimental
trace in the forward tuning (yellow curve) followed by one trace in
the backward tuning (white curve) with successive transitions of multiple-soliton
states from ${N=7}$ to ${N=0}$ (no solitons); (h-j) frequency comb
spectra in soliton states with ${N=1,2,3}$, measured during the backward
tuning in a $100~\mathrm{GHz}~\mathrm{Si_{3}N_{4}}$ microresonator;\label{fig_1}}
\end{figure*}

In this paper we report the discovery of a phenomenon that allows to induce deterministically transitions to states with less solitons (i.e. from $N$ to $N-1$), and thereby to reliably reach the single soliton state. The phenomenon is not explained by standard theoretical simulations based on the Lugiato-Lefever equation (LLE) or coupled mode equations models \cite{chembo2010modalexpansionapproach,Chembo2013LLE,Coen2013modelling,godey2014stabilityLLE,herr2014soliton,Lamont2013stabil}.
We present detailed analysis of the observed phenomenon, in two
microresonator platforms where the thermal locking is possible, and demonstrate its universal nature. The reported findings allow to switch between multiple-soliton states by sequentially reducing the number ${N}$ of initially created
solitons (with a routine simple enough to be carried out by a micro-controller),
to monitor and control the switching, and to hold the targeted soliton
state, preventing it from decay. 
Especially, the single soliton state can be deterministically and reliably induced, which is imperative to a wide range of
applications. 
The presented results contribute to the physical understanding of switching behavior of the DKS, highlight the influence of thermal effects and provide a rich toolbox for the study of the multiple-soliton dynamics. From an applied perspective, the results present a route to making reliable pulse sources and frequency combs based on DKS at microwave repetition rates in optical microresonators.

\section*{Results}

\label{Backward tuning}

\noindent \textbf{Observation of switching between dissipative Kerr
soliton states by laser backward tuning.} The principles of microresonator
frequency comb generation and the formation of dissipative Kerr solitons
(DKS) are shown in Fig.\ref{fig_1}(a). CW laser light is coupled to a high-$Q$
optical resonator, where modulation instability (MI) and cascaded four-wave-mixing
processes lead to the formation of a broadband frequency comb. In
this work we study two microresonator platforms: ${\mathrm{Si_{3}N_{4}}}$
on-chip ring microresonators \cite{Levy2010compatible,moss2013cmos,Foster2011SINcomb,Brasch2014Cherenkov}
(Fig.\ref{fig_1}(b)) and ${\mathrm{MgF_{2}}}$ crystalline resonators \cite{Ilchenko2004crystal,Grudinin2006highq,Liang11NIRcomb,herr2014soliton}
(Fig.\ref{fig_1}(c)). The laser tuning technique was developed as an effective method to the formation of dissipative Kerr solitons \cite{herr2014soliton}, in
which the CW pump laser is tuned (from short to long wavelengths)
over the cavity resonance, referred to as the ``forward tuning''.
Initially, the CW pump is in the blue-detuned regime. The cavity resonance
is shifted due to the slow thermal and fast Kerr nonlinearity of the microresonator,
resulting in a self-locking of the cavity resonance to the pump laser
\cite{Braginsky1989triangle,carmon2004thermalstability}. In this
regime the Kerr comb formation can be observed. The mechanism results in a triangular trace in
the generated comb light, over the pump frequency detuning. When the
pump is tuned over the cavity resonance, it enters the effectively
red-detuned regime where multiple dissipative Kerr solitons (i.e.
multiple-solitons) can be formed. The soliton state is accompanied with
a step-like power trace in the generated comb light, where the step
height corresponds to the number of solitons ($N$) inside the resonator.
Transitions to states with lower number of solitons may also occur
and the power trace will exhibit a characteristic steps \cite{herr2014soliton}.
Eventually, by stopping the pump laser tuning at a step while ensuring the thermal equlibrium in the resonator, stable multiple-soliton
and even single soliton states can be accessed (Fig.\ref{fig_1}(d)). This forward tuning
method was applied in ${\rm MgF_{2}}$, ${\rm Si_{3}N_{4}}$
and silica resonators for single dissipative Kerr soliton generation
\cite{herr2014soliton,Brasch2014Cherenkov,Yi2015solitoncaltech}.

However, in on-chip microresonators, the thermal
nonlinearity significantly impacts the soliton step pattern, such
that single soliton states become rarely accessible with the forward
tuning method. Figure \ref{fig_1}(e) shows 200 overlaid experimental power
traces of the generated comb light obtained in a ${\rm Si_{3}N_{4}}$
microresonator in the forward tuning, in which only multiple-soliton
states are stochastically accessed having ${N=6}$ (predominantly), 7, 8 or 9. 
Careful studies further reveal several common features in
${\rm Si_{3}N_{4}}$ microresonators, irrespective of the employed
pump power (see Supplementary Information (SI) for more details) and
the laser tuning speed: (1) the distributions mostly consist of the
traces with one step corresponding to high-${N}$ multiple-soliton
state; (2) the accessible step length decreases with decreasing $N$;
(3) the number of generated solitons increases with increasing pump
power. All of these imply that the single soliton state is not readily
accessible in the forward tuning.

Remarkably, an additional laser tuning towards shorter wavelengths (``backward tuning'') provides a way to reliably access the single soliton state starting from an arbitrary multiple-soliton state.
 The result of this backward tuning sequence, shown in Fig.\ref{fig_1}(f),
allows for successive extinction of intracavity solitons (soliton
switching) down to the single soliton state (${N\! \to\! N\!-\!1\! \to\! \dots \!\to\! 1}$).
Figure \ref{fig_1}(g) shows \textit{\emph{one}} trace of the generated light
of the ${\rm Si_{3}N_{4}}$ microresonator, where switching from seven
solitons to the single soliton is observed. Strikingly, the power trace
of the generated comb light reveals a \textit{regular staircase} pattern
with equal stair length and height. The exact soliton
number in each step can be precisely inferred from the step height.
The pattern is almost identical over multiple experimental runs (using
the same tuning speed and pump power) \textit{regardless} of the initial
soliton number ${N}$. Each transition between multiple-soliton states occurs with the extinction of preferably \emph{one} soliton at a time, which is confirmed by the relative positions of the intracavity solitons that are retrieved from the optical spectrum (cf. insets in Fig.\ref{fig_1}(h-j)).

\begin{figure*}
\includegraphics[width = 2 \columnwidth]{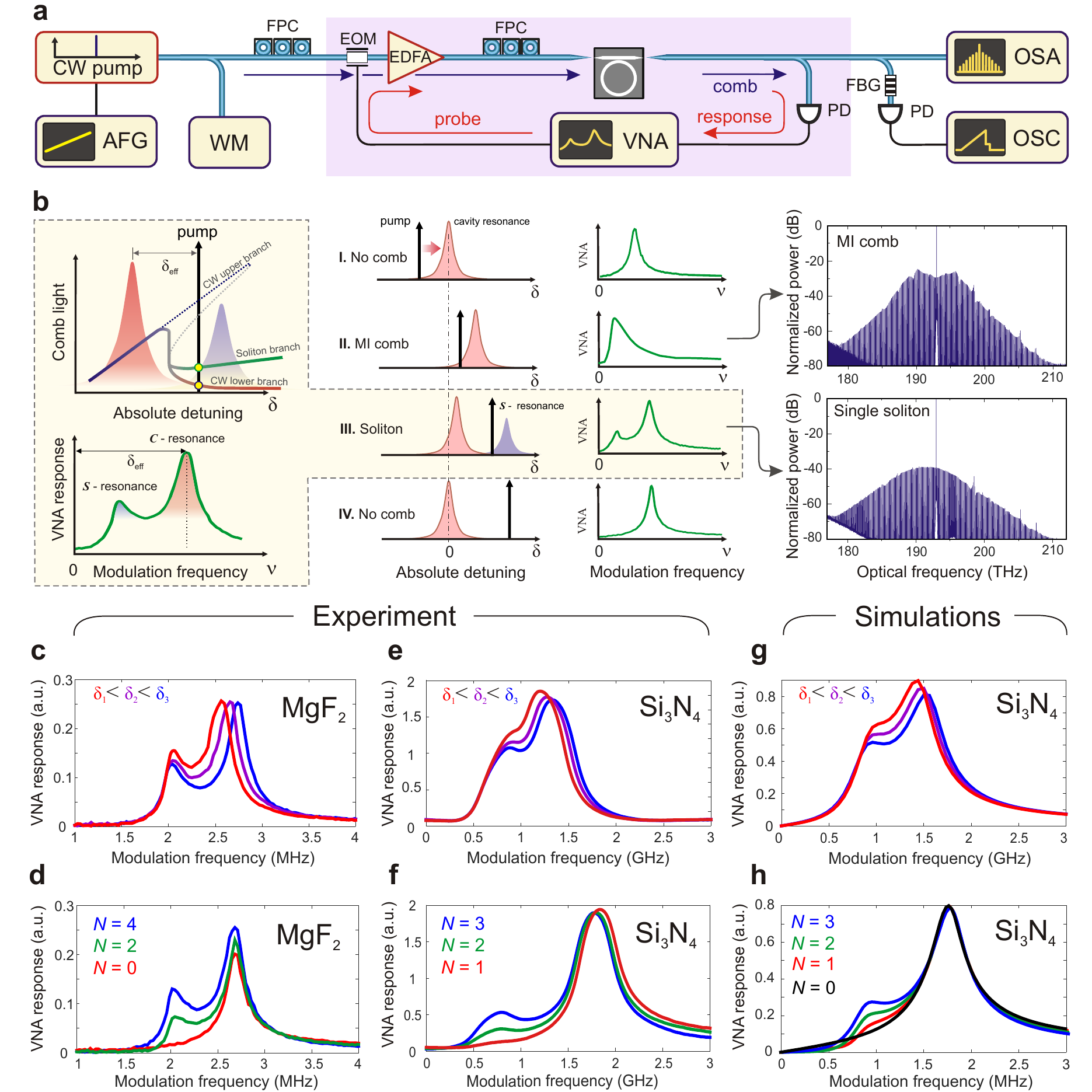}
\protect\caption{\textbf{Dynamical probing of dissipative temporal Kerr solitons (DKS)
in microresonators.} (a) Setup scheme used for soliton generation,
non-destructive soliton probing and deterministic soliton switching.
An external cavity diode laser (CW pump) is used as a pump source.
AFG, arbitrary function generator; EDFA, erbium doped amplifier; FPC,
fiber polarization controller; WM, wavelength meter; VNA, vector network
analyzer; OSA, optical spectrum analyzer; OSC, oscilloscope; PD, photodiode; EOM, electro-optical phase modulator;
PM, phase modulator; FBG, fiber Bragg grating. (b) (Left, top) Diagram of the double-resonance modulation response in the soliton state. The green line indicates the power trace of the soliton component which is evolved from the high-intensity branch of the bistability (blue line). The pump is tuned in the bistability range (in the effective-red detuned regime). Therefore, both the soliton branch and the low-intensity continuous (CW) branch (red line) are supported in the system, each corresponds to a resonance, i.e. the $\mathcal{S}$-resonance and the $\mathcal{C}$-resonance. (Left, down) Double-resonance modulation response from VNA. The high-frequency peak indicates the $\mathcal{C}$resonance and the low-frequency is the $\mathcal{S}$-resonance. (Middle) Four
stages of the microresonator frequency comb formation and corresponding
VNA modulation response when the pump laser is forward tuned over
the resonance: (I) No comb, the pump is blue-detuned; (II) Chaotic MI comb
state; (III) Soliton state; (IV) No comb, the pump is red-detuned.
(right) Frequency comb spectra corresponding to the chaotic MI operation regime and the single soliton state. (c-f) Experimental
double-resonance response of various multiple-soliton states at different
detunings for ${\mathrm{MgF_{2}}}$ and ${\mathrm{{\rm Si_{3}N_{4}}}}$
microresonators; (i-j) Simulated double-resonance response for ${\rm Si_{3}N_{4}}$
microresonator used in the work.\label{fig_2}}
\end{figure*}

In experiments, the backward tuning process must be adiabatic to induce
the successive reduction of the soliton number: the thermal equilibrium
is required at each multiple-soliton state. This prerequisite is satisfied
by choosing a tuning speed much slower than the thermal relaxation
rate that depends on the effective mode volume and the thermal diffusivity of a
microresonator \cite{fomin2005nonstationary}. For the employed ${\rm Si_{3}N_{4}}$
microresonator the backward tuning speed is chosen ${\sim40~\mathrm{MHz/s}}$, while the forward tuning speed is ${\sim100~\mathrm{GHz/s}}$. In this
way all soliton states (${\le N}$) are deterministically accessible.
In contrast to the robust backward tuning that enables successive
extinction of intracavity solitons, the forward tuning in $\mathrm{Si_{3}N_{4}}$ microresonators
always leads to collective extinction of solitons.

The backward tuning was also studied in ${\mathrm{MgF}_{2}}$ crystalline
microresonators, where the successive soliton switching to the single
soliton state is also achieved. In contrast to the ${\rm Si_{3}N_{4}}$ platform,
the single soliton state can directly be accessed with the forward
tuning in ${\mathrm{MgF}_{2}}$ microresonators \cite{herr2014soliton}.
Yet, this requires fine adjustments on the coupling, the pump power
and the tuning speed. The backward tuning, on the
other hand, is much more robust and significantly facilitates the
generation of single soliton states for crystalline resonators.

The soliton switching in both $\mathrm{Si_{3}N_{4}}$ and crystalline
$\mathrm{MgF}_{2}$ resonator, proves that the backward tuning represents a universal approach to the generation of a single soliton state in microresonators, provided that the thermal locking can be achieved.

\label{Soliton probing} \textbf{Non-destructive probing of the soliton
response.} Dissipative Kerr solitons in microresonators represent
stable and self-reinforcing intracavity light patterns resulting from
double balance between pump and cavity losses, as well as chromatic
dispersion and Kerr nonlinearity of the resonator. 
The key parameter of such soliton state is the effective laser frequency detuning that determines both the amplitude and the duration of soliton pulses \cite{herr2014soliton}. This detuning is defined as ${2\pi\delta_{\mathrm{eff}} = \widetilde\omega_{\rm 0}-\omega_{\rm p}}$,
where ${\widetilde\omega_{\rm 0}}$ indicates the frequency of a cavity resonance and ${\omega_{\rm p}}$ is the pump laser frequency.
In experiments the pump frequency is precisely controlled, but the
resonance frequency is thermally shifted from the initial cold cavity resonance
frequency ${\omega_{\rm 0}}$, making it \textit{a priori }not possible
to evaluate the effective detuning. 
On the other hand, the absolute detuning $2\pi\delta = \omega_{\rm 0}-\omega_{\rm p}$ can be introduced and measured as the position of the pump frequency relative to the fixed cold cavity resonance.
It has been shown that solitons are supported within a certain range of the effective detuning \cite{herr2014soliton,karpov2015raman},
when the pump is effectively-red detuned (${\omega_{\rm p} < \widetilde\omega_{\rm 0}}$),
which we refer to as the soliton existence range for a given constant
input power.

We developed a non-destructive soliton probing scheme that allows
to track the effective detuning and extract the soliton number $N$ of microresonator frequency combs. The setup, presented
in Fig.\ref{fig_2}(a), employs a pump laser, whose frequency is phase modulated
using a vector network analyzer (VNA), that produces weak optical
sidebands with sweeping frequency ($\nu$) in the range $5~ \mathrm{kHz}
- 4.5~\mathrm{GHz}$, which probe the state of the microresonator
system. The complex modulation response to such probes is measured
by the VNA.

This probing method enables to identify different stages
in the generation of frequency comb, including the soliton formation,
see Fig.\ref{fig_2}(b). First, when the pump is in the blue-detuned regime
(${\omega_{\rm p} > \widetilde\omega_{\rm 0}}$), away from the cavity resonance, the modulation
response on the VNA shows a Lorenzian-like resonance profile that
corresponds to the cavity resonance with the peak position indicating
$|\delta_{\mathrm{eff}}|$. Second, when (forward) tuning the pump frequency
into the cavity resonance, where the frequency comb in the chaotic MI regime
is observed, the modulation response shows an asymmetric profile with
the peak position being fixed, indicating the thermal and Kerr locking
of the cavity resonance to the pump frequency. Third, when the frequency
comb is in the soliton state, with the pump laser tuned in the soliton
existence range in the red-detuned regime,
the modulation response shows unexpectedly a \emph{double-resonance}
feature. Finally, when the pump frequency is tuned out of the soliton
existence range where no comb is observed, the modulation response
shows again a single, Lorenzian-like resonance similar to the first
stage.

\begin{figure*}
\includegraphics[width = 2 \columnwidth]{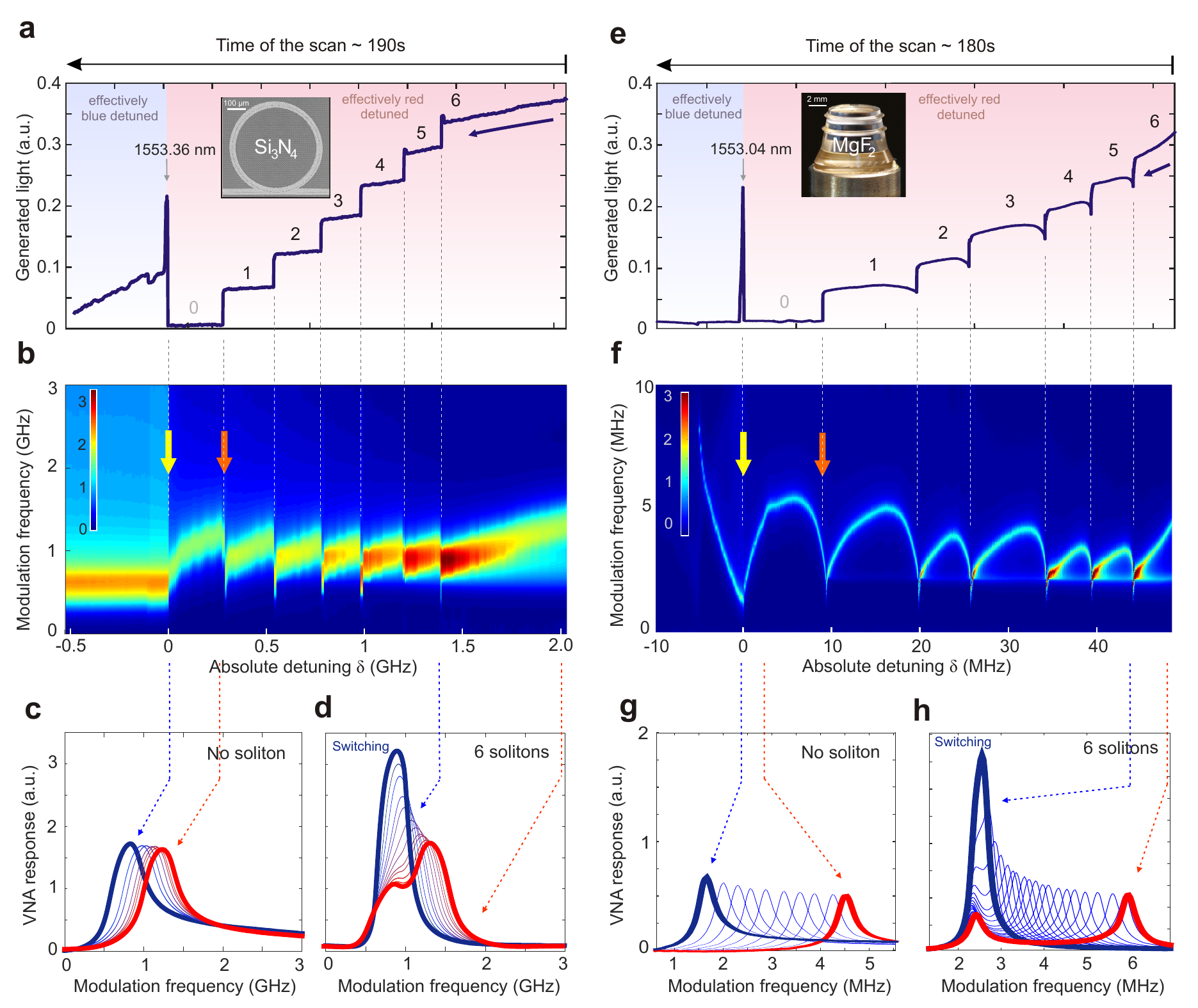}
\protect\caption{\textbf{Deterministic switching of the soliton states.} (a) The power
trace of the generated light obtained from $100$ $\mathrm{GHz}$
$\mathrm{Si_{3}N_{4}}$ microresonator with the backward pump tuning
from multiple-soliton with $N=6$ (effectively red detuned) to the effectively
blue detuned regime; (b) Set of $500$ concatenated VNA traces that
were taken during the backward tuning shown in (a). The orange arrow
indicates the transition from a single soliton state to no-soliton
state, while the pump is still red detuned with respect to the cavity
resonance. The yellow arrow indicates the transition from the red
detuned operating regime to the blue detuned regime; (c) Evolution
of the modulation response during the backward tuning process in the
effectively red detuned regime, with no soliton presented ($N=0$);
(d) Evolution of the modulation response in the multiple-soliton state
with $N=6$; (e) The power trace of the generated light obtained from
$14~\mathrm{GHz}~\mathrm{MgF_{2}}$ crystalline resonator with
the backward pump tuning from multiple-soliton state with $N=6$ (effectively
red detuned) to the effectively blue detuned regime; (f) Set of $\sim1700$
concatenated VNA traces that were taken during the backward tuning
shown in (e); (g) Evolution of the modulation response during backward
tuning in the state with no soliton presented; (h) Evolution of modulation
response in the multiple-soliton state with $N=6$.\label{fig_3}}
\end{figure*}

The double-resonance response that is observed in the presence of soliton states can be attributed to the superposition of weak continuous background and intense soliton pulses \cite{herr2014soliton}. Due to different intensities each component induces a different Kerr shift to the cavity resonance which we can discriminate by the modulation probing. Since the pump is far detuned from the cavity resonance, the high-frequency peak in the modulation response corresponds to the cavity resonance that is slightly shifted by the CW component ($\mathcal{C}$-resonance). The peak position in this way indicates the effective detuning. On the other hand, the resonance shifted by solitons appears as the low-frequency peak ($\mathrm{\mathcal{S}}$-resonance). The position of the $\mathcal{S}$-resonance is nearly fixed as it depends on the intensity of individual soliton, while the magnitude is related to the number of solitons ($N$).

We applied the non-destructive soliton probing to both $\mathrm{Si_{3}N_{4}}$ and ${\rm MgF_{2}}$ microresonators. The double-resonance response is observed in both platforms when having soliton state frequency combs, and is investigated with different soliton number $N$ and pump detunings, see Fig.\ref{fig_2}(c-f). The response is qualitatively similar for both platforms. The peak position of the $\mathcal{C}$-resonance varies with the pump frequency (Fig.\ref{fig_2}(c,e)), while the $\mathrm{\mathcal{S}}$-resonance frequency is practically
fixed as predicted. The peak height of the $\mathrm{\mathcal{S}}$-resonance linearly depends on the soliton number
$N$ (Fig.\ref{fig_2}(d, f)). We also performed a theoretical analysis of the non-destructive soliton probing scheme, which confirms the double-resonance response of a soliton state (see Fig.\ref{fig_2}(g, h)). 

The response of dissipative Kerr solitons to weak amplitude pump modulation was earlier numerically investigated  in \cite{matsko2015feshbach}. While two peaks in the response were also numerically observed in that work (and attributed conceptually to Feshbach and relaxation oscillations in the presence of third order dispersion), the present work reveals the underlying physical origin of the soliton probing scheme, not requiring higher order dispersion. Moreover, phase modulation provides higher contrast of the modulation response.

As a way to extract the effective pump detuning $\delta_{{\rm eff}}$,
the probing technique enables to precisely track the process of microresonator
frequency comb generation in experiments. In a soliton state, thermal
drifts of the cavity resonance originating from various external sources
may cause variations of $\delta_{{\rm eff}}$. Based on the modulation
response, the effective detuning can be monitored and adjusted (e.g.
by tuning the pump frequency) in order to maintain the soliton state
within the soliton existence range. In practice, feedback-locking
of $\delta_{{\rm eff}}$ is possible, which allows for long-term operation
of a soliton state in a microresonator.

\label{Deterministic switching} \textbf{Deterministic switching of
soliton states.} We next investigate the transitions of soliton states
in the laser backward tuning by applying the non-destructive
soliton probing in $\mathrm{Si_{3}N_{4}}$ microresonators. We first
employ forward tuning in order to generate a multiple-soliton state with
$N=6$, and then we perform the slow backward tuning. The power trace
of the generated light in the microresonator again shows the staircase
pattern in the backward tuning, which corresponds to successive soliton
switching from $N=6$ to the single soliton state (Fig.\ref{fig_3}(a)).
The VNA traces are simultaneously recorded and continuously stacked
in order to monitor the evolution of the modulation response during
the process (see Fig.\ref{fig_3}(b)).

The experiments reveal a relationship between the evolution of modulation
response and the soliton switching. Within each soliton step, the
$\mathcal{C}$-resonance shifts towards the $\mathcal{S}$-resonance due to the
decrease of the effective detuning when the laser is tuned backward.
When the two resonances overlap, the amplitude of $\mathcal{S}$-resonance
is significantly enhanced, leading to a high-intensity single-peak
profile (Fig.\ref{fig_3}(d)). The phenomenon is also confirmed by the theory
(cf. SI). The next moment after having such a response, soliton switching
occurs, which results in the power drop in the generated light trace
as one soliton is extinct (${N\! \to\! N\!-\!1}$). After the switching,
the $\mathcal{C}$-resonance abruptly separates from the $\mathcal{S}$-resonance.
Meantime, while still being Kerr locked, the $\mathcal{S}$-resonance
intensity is reduced to a lower level than the previous state, since
the number of solitons is reduced by one. In the absence of solitons
(${N=0}$), the $\mathcal{S}$-resonance equally is absent in the
modulation response, but the $\mathcal{C}$-resonance is still present and captured
(Fig.\ref{fig_2}(c)) .

The same measurement was carried out in ${\rm MgF_{2}}$ resonators,
see Fig.\ref{fig_3}(e-h). Similar switching dynamics as in $\mathcal{\mathcal{\mathrm{Si_{3}N_{4}}}}$
microresonators are observed: (1) the power trace shows staircase
profile of successive soliton switching; (2) the backward tuning shifts
the VNA $\mathcal{C}$-resonance towards the $\mathcal{S}$-resonance; (3) soliton
switching occurs with the overlap of $\mathcal{C}$- and $\mathcal{S}$-resonances
and the enhancement of the $\mathcal{S}$-resonance intensity. However, there are
several details which differ between $\mathcal{\mathcal{\mathrm{Si_{3}N_{4}}}}$
and ${\rm MgF_{2}}$ platforms. First, the optical quality factor
$Q$ of ${\rm MgF_{2}}$ crystalline resonators (${\sim10^{9}}$)
is three order of magnitude higher than for $\mathcal{\mathcal{\mathrm{Si_{3}N_{4}}}}$
micro-rings (${\sim10^{6}}$). The $\mathcal{C}$- and $\mathcal{S}$-resonances
in the modulation response of crystalline resonator are better resolved
as a result of the narrower linewidth. The laser tuning range in $\mathcal{\mathcal{\mathrm{Si_{3}N_{4}}}}$
microresonators is ${\mathcal{O}(1{\rm {~GHz})}}$, while that in
$\mathrm{MgF}_{2}$ resonators is ${\mathcal{O}(1{\rm {~MHz})}}$.
Second, after each soliton switching the $\mathrm{MgF}_{2}$ resonator
shows slower recoil of the $\mathcal{C}$-resonance than the $\mathcal{\mathcal{\mathrm{Si_{3}N_{4}}}}$
microresonator. This is attributed to the distinct thermal relaxation
of the two platforms. The $\mathrm{MgF}_{2}$ resonator has a larger
effective mode volume and physical size than the chip-scale $\mathcal{\mathcal{\mathrm{Si_{3}N_{4}}}}$
micro-ring resonators such that the thermal relaxation time is longer.
In the evolution of the modulation response of the $\mathrm{MgF}_{2}$
resonator (Fig.\ref{fig_3}(f)), the recoil of the $\mathcal{C}$-resonance leaves curved
trajectory while it is very abrupt in the $\mathrm{Si_{3}N_{4}}$ microresonator (Fig.
\ref{fig_3}(b)).

The non-destructive soliton probing scheme combined with the backward
tuning allows an understanding of the soliton switching dynamics in
microresonators. The modulation response clearly predicts the switching
and therefore provides a convenient tool to control the soliton states
and induce switching \textit{on demand}. In experiments, one can perform
deterministic switching by tuning the pump frequency, while monitoring
the effective laser frequency detuning revealed by the VNA response.

\begin{figure*}
\includegraphics[width = 2 \columnwidth]{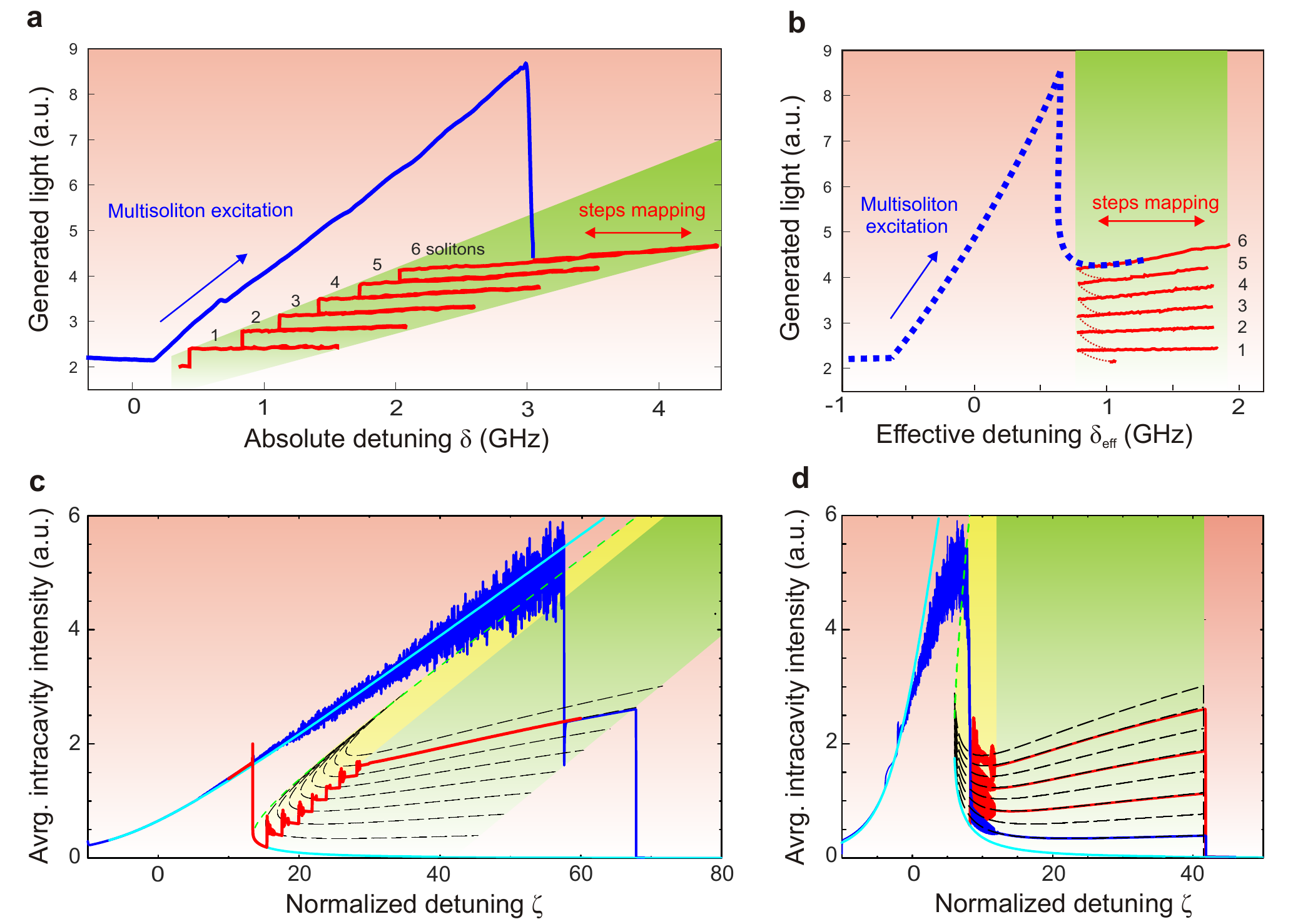}
\protect\caption{\textbf{Experimental mapping of soliton existence range and numerical
simulations}. (a) Experimental measurements of the generated comb
light with respect to the absolute detuning. The blue
curve shows the trace in the forward tuning. The red curve indicates
the entire soliton existence range. Zero absolute detuning corresponds to $\sim 1553.4\mathrm{nm}$. (b) Experimental trace from (a)
plotted in terms of the effective detuning measured from the modulation
response with the VNA. Hypothetical trace of forward tuning is shown
in blue dashed line, because the effective detuning in this process
can not be reliably measured with the VNA. (c, d) Numerical simulations
and analytical solutions of the backward tuning in $\mathrm{Si_{3}N_{4}}$
with (c) and without (d) thermal effects. Normalized detuning used in the simulation: $\zeta = 2(\omega_{\rm 0}-\omega_{\rm p})/\kappa$, where $\omega_{\rm 0}$ is the resonance frequency, $\omega_{\rm p}$ is the pump frequency and $\kappa$ is the resonance linewidth. Blue lines indicate initial
excitation of a multiple-soliton state in the forward tuning. Red
lines indicate the backward tuning. Light blue lines indicate the
stable branch of the nonlinear induced tilted resonance (in the CW
mode). Dashed green lines indicate the unstable branch. The yellow
area allows for the formation of breathing soliton states. The green
area indicates the formation of stable solitons. Solitons cannot exist
in red area. The dashed lines show analytical description of soliton
steps, with analytical solution of soliton states in the system.\label{fig_4}}
\end{figure*}

\textbf{Thermally enabled transitions of soliton states.} We attribute
the successive soliton switching in backward tuning to the thermal
nonlinearity of optical microresonators. Due to material absorption,
the intracavity energy of a soliton state thermally shifts the cavity
resonance via thermal expansion and thermal change of the refractive index: $\widetilde\omega_{0}=\omega_{0}-\Delta_{\rm T}$, where
${\Delta_{\rm T}}$ is the thermally induced resonance shift which is approximately (neglecting cross term)
proportional to the energy of intracavity field:
\begin{equation}
\Delta_{{\rm T}}(N)\propto E_{{\cal C}}+N\cdot E_{{\cal S}}
\end{equation}
where ${E_{{\mathcal C}}}$ is the energy of the ${\mathcal C}$ component, ${E_{{\cal S}}}$
is the energy of one soliton and $N$ the number of solitons.
Thus, the effective detuning can be expressed as ${2\pi\delta_{{\rm eff}} = \omega_{\rm 0}-\Delta_{\rm T}-\omega_{\rm p}}$.
Physically, the soliton switching occurs when the laser backward tuning reduces
the detuning to the bifurcation point of the system.
This boundary value can be identified from the
VNA trace and is represented by the position of the $\mathcal{S}$-resonance.
After the switching, one soliton is extinct which decreases the
energy in the cavity, and thereby reduces the thermal shift ${\Delta_{\rm T}}$.
This spontaneously stabilizes the system in a new soliton state, by
effectively increasing the effective detuning. The process is
reflected in the evolution of the modulation response (see Fig.\ref{fig_3}(b))
as a separation of $\mathcal{C}$- and ${\mathcal{S}}$-resonance after
the switching. It should be also noted that the recoiled $\mathcal{C}$-resonance
frequency is similar after each switching event, because the resonator
loses approximately similar amount of energy. Overall, the thermal
nonlinearity lifts the degeneracy of soliton states with respect to the pump frequency.

\textbf{Full mapping of the soliton induced multi-stability diagram
in optical microresonator.}
The pump backward tuning enables deterministic
and successive soliton switching, opening access to soliton
states ${N,N-1,...,1}$. It is therefore possible to experimentally
explore the soliton existence range in terms of the absolute and the effective detuning in each state, which to authors best knowledge has never been directly experimentally measured for cavity solitons of any kind.
In terms of effective detuning, we express the soliton existence range as ${\delta_{{\rm s}}<\delta_{{\rm eff}} <\delta_{{\rm max}}}$.
The lower boundary $\delta_{{\rm s}}$ is identified in the backward soliton
switching: it corresponds to the frequency where the $\mathcal{C}$-resonance
and the fixed $\mathcal{S}$-resonance overlap. In the studied $\mathrm{Si_{3}N_{4}}$
microresonator under chosen pumping conditions this quantity is measured as ${\delta_{\rm s}\sim 0.78{\rm {~GHz}}}$.
The upper detuning boundary $\delta_{{\rm max}}$ of the soliton existence range can be explored for each soliton state when the pump laser is tuned forward until the soliton comb disappears. 
Based on the theory and standard LLE simulations, this detuning is expected to be identical for all states corresponding to different number of solitons \cite{herr2014soliton}(see Fig.\ref{fig_4}(d) and also SI), as the boundary of the energy balance of dissipative Kerr solitons.
In experiments under the same pumping conditions such maximum effective
detuning ${\delta_{\rm max}}$ is found for all soliton states as ${\sim 2.0{\rm {~GHz}}}$, yet no clear feature in the modulation response enables to predict this maximum boundary.

Figure 4(a) displays a \textit{one-trace} mapping of six steps of soliton
states in $\mathrm{Si_{3}N_{4}}$ microresonator as a function of the
absolute pump frequency (wavelength) (i.e. the absolute detuning ${\delta}$). For each soliton step, we first
tune the pump forward approaching the maximum detuning ($\delta \! \to\!\delta_{{\rm max}}$),
and then tune backward towards the soliton switching point (${\delta\! \to\!\delta_{{\rm s}}}$)
where the soliton state is switched from $N$ to $N-1$.
Since the thermally induced cavity resonance shift is included in
the absolute frequency detuning, we observe that the soliton existence
range in the absolute detuning is increasingly offset for a larger
number of soliton. This creates the staircase pattern of the generated
light and enables successive soliton switching. However, if the generated
light trace is plotted with respect to the \emph{effective} laser detuning
(${\delta_{{\rm eff}}}$) as done in Fig.\ref{fig_4}(b), it appears that all
the soliton steps are stacked vertically within the range ${\delta_{{\rm s}}<\delta_{{\rm eff}} <\delta_{{\rm max}}}$,
which corresponds to the expected theoretical diagram when the thermal
effect is neglected \cite{herr2014soliton}.

We performed numerical simulations based on both LLE and coupled mode equations with the additional thermal relaxation equation included (cf. SI) which verify that the
deterministic soliton switching is enabled by the thermal nonlinearity of the microresonator
(Fig.\ref{fig_4}(c, d)). By including the thermal effects into numerical simulations, we are able to reproduce the staircase power trace, corresponding to the successive reduction of the soliton number in the backward pump tuning (cf. red curve in Fig.\ref{fig_4}(c)). Analytical power traces of soliton steps (black dashed lines) indicate soliton existence
ranges for multiple-soliton states with different $N$. They
reveal a displacement of the soliton existence range between different
soliton states (qualitatively similar to the measured in Fig.\ref{fig_4}(a)) as a consequence of the thermal nonlinearity.

When the thermal effects in simulations are ``switched off'', soliton
steps are well aligned and the soliton existence range is again degenerate
with respect to the soliton number ($N$), see Fig.\ref{fig_4}(d). No soliton
switching is therefore observed in the backward tuning. Numerical
simulation also revealed the soliton breather states that is considered as a intermediate state between the chaotic MI operation regime and the stable
soliton state. In the breather state, the soliton pulse peak power
and the pulse duration, as well as the average intacavity energy, will experience periodical oscillations.
This induces thermal perturbations to the cavity resonance and initiates the soliton switching.

\section*{Discussion}

\label{Discussion}

We experimentally, numerically and analytically demonstrate the discovery
that soliton states in a microresonator are not detuning degenerate, and can be individually addressed by laser detuning.
We demonstrate that this
effect is platform independent and can be used in a 
laser backward tuning process to achieve a successive reduction of the soliton
number (${N\! \to\! N\!-\!1\! \to\! \dots \!\to\! 1}$). This deterministic switching is enabled by the thermal nonlinearity of the microresonator and provides a route to obtain
a single soliton state from an arbitrary multiple-soliton state. We have
shown non-destructive soliton probing technique, which enables to
track the thermal impact of external perturbations of the system on
its stability. The technique also allows to lock the soliton state
against the impact of these perturbations and gives a clear insights
of soliton dynamics inside the cavity. Combining this technique with
the laser backward tuning allows for deterministic soliton switching
and makes accessible any target multiple-soliton state in a predictable
way. The results are in good agreement with analytical treatment of
the soliton comb including thermal effects as well as numerical simulations, and can be applied to all Kerr nonlinear microresonators.

\section*{Methods}

\label{Methods} $\mathcal{\mathcal{\mathrm{Si_{3}N_{4}}}}$ microresonators
investigated in this work were fabricated using the \textit{Photonic Damascene
process} \cite{Pfeiffer2015Damascene}. Microresonators have FSR of
$100$ $\mathrm{GHz}$. A single mode ``filtering'' section was added to the micro-rings in order to suppress high-order modes \cite{kordts2015higher}. The dispersion parameters of the microresonators are measured using the frequency comb assisted laser spectroscopy method \cite{Delhaye2009disp}: ${\frac{D_{2}}{2\pi}=1-2~{\rm MHz}}$,
${\frac{D_{3}}{2\pi}=\mathcal{O}(1~{\rm kHz})}$ (where the resonance frequencies near ${\omega_0}$ are expressed in a series ${\omega_\mu = \omega_0 + \sum_{i\ge 1}  D_{i}\mu^{i}/{i!}}$, where ${i \in \mathbb{N}}$, ${\mu \in {\mathbb Z}}$ is the mode number). Pumped resonance
is at $1553.4~\mathrm{nm}$. Tuning speed for soft excitation is ${\sim}$
$1$ $\mathrm{nm/s}$. Pump power is ${\sim}$ $2-3$ $\mathrm{W}$ on
a chip.

The detailed scheme of the experimental setup is presented in the Figure 2(a).
The ${\mathrm{Si_{3}N_{4}}}$ resonator is pumped
with a CW laser light from an external-cavity diode laser amplified by an erbium-doped fiber amplifier
(EDFA) to $3-5$ $\mathrm{W}$. The CW pump is coupled to the on-chip
resonator using lensed fibers with coupling losses of $2.5-3~\mathrm{dB}$
per facet. For the soliton probing measurements $10 ~\mathrm{GHz}$
electro-optical phase modulator (EOM) is placed before EDFA with additional polarization controller
for adjusting input polarization. The pump frequency wavelength in
the pump backward tuning is measured by a wave-meter with resolution
of ${\sim50{\rm {~MHz}}}$. For the long sweeps an arbitrary function
generator is used. The output signal from the chip is split in several
paths among OSA (for the measurements of combs spectra), oscilloscope
(for the measurements of generated light by filtering out the pump
with FBG) and a VNA receiver (for the measurements of and modulation
response).

The $\mathrm{MgF_{2}}$ crystalline resonator was fabricated by diamond
turning of a cylinder blank and subsequent hand polishing to achieve
high $Q$ (linewidth $\frac{\kappa}{2\pi}=100~{\rm kHz}$). The diameter
of 5 mm yields a FSR $\frac{D_{1}}{2\pi}=14$ GHz. The dispersion
parameters at the pump wavelength of $1553$ $\mathrm{nm}$ are: ${\frac{D_{2}}{2\pi}=1.9~{\rm kHz}}$,
${\frac{D_{3}}{2\pi}=\mathcal{O}(1~{\rm Hz})}$. The pump laser (fiber
laser, wavelength $1553$ $\mathrm{nm}$; short-term linewidth $10$
$\mathrm{kHz}$) is amplified to $\sim250~{\rm mW}$. The relative
laser frequency is monitored by counting the heterodyne beat between
the pump laser and a reference laser stabilized to an ultra-stable
cavity. The light is evanescently coupled to a WGM with a tapered
optical fiber.

%

\section*{Acknowledgements}

\label{Acknowledgements}

\begin{acknowledgments}
\emph{Acknowledgements}. This publication was supported by Swiss National
Science Foundation (SNF) as well as Contract
W31P4Q-14-C-0050 from the Defense Advanced Research Projects Agency
(DARPA), Defense Sciences Office (DSO). This
material is based upon work supported by the Air Force Office of Scientific Research,
Air Force Material Command, USAF under Award No. FA9550-15-1-0099. This publication was supported by funding from the European Space Agency (ESA), European Space Research and Technology Centre (ESTEC). G.L., V.E.L. and M.L.G. were supported 
by the Ministry of Education and Science of the Russian Federation project 
$\#$4.585.21.0005.  The authors gratefully acknowledge valuable discussions with Michael Geiselmann, Martin H.P. Pfeiffer, John D. Jost and Victor Brasch. All samples were fabricated and grown in the Center of MicroNanoTechnology (CMi) at EPFL.
\end{acknowledgments}

\section*{Authors contributions}
M.K. designed and performed the experiments and analyzed the data. H.G. conceived and initiated the numerical simulations. E.L. performed experiments in $\mathrm{MgF}_{2}$ microresonators and analyzed the data. A.K. fabricated the ${\rm Si_{3}N_{4}}$ samples and M.P. developed the fabrication method. G.L. and V.E.L. assisted in simulations. M.L.G. developed the theory and performed the simulations. M.K., H.G., E.L., M.L.G., T.J.K. discussed all data in the manuscript. M.K. and H.G. wrote the manuscript, with input of E.L., M.L.G., T.J.K. T.J.K. supervised the project. 
\label{Authors contributions}

\end{document}